\documentclass{ws-ijmpd-arXiv160403065}

\usepackage{amsmath,amssymb,revsymb,graphicx,dcolumn}

\newcommand{\beq}{\begin{equation}}
\newcommand{\eeq}{\end{equation}}
\newcommand{\beqa}{\begin{eqnarray}}
\newcommand{\eeqa}{\end{eqnarray}}
\newcommand{\bsubeqs}{\begin{subequations}}
\newcommand{\esubeqs}{\end{subequations}}
\newcommand{\half}{{\textstyle\frac{1}{2}}}

\begin{document}

\markboth{F.R. Klinkhamer}
{A generalization of unimodular gravity ...  }

%
\catchline{}{}{}{}{}
%

\title{A GENERALIZATION OF UNIMODULAR GRAVITY
      WITH VACUUM-MATTER ENERGY EXCHANGE}

\author{F.R. KLINKHAMER}
\address{Institute for
Theoretical Physics, Karlsruhe Institute of Technology (KIT),\\
76128 Karlsruhe, Germany\\
frans.klinkhamer@kit.edu}

\maketitle


\begin{abstract}
An effective theory of gravity in the infrared is proposed,
which involves the determinant of the metric relative to
the determinant of a prior metric taken to be that of
Minkowski spacetime.
This effective theory can be interpreted as a generalization
of unimodular gravity.
In a cosmological context with ultrarelativistic or cold matter,
the resulting field equations
have only one solution, empty Minkowski spacetime
(selected by the prior metric of the theory).
The introduction of energy exchange between vacuum and matter
gives rise to nonstatic cosmic solutions.
It is found that Minkowski spacetime (from the prior metric)
appears as an attractor of the dynamic equations.
A further result is that energy-momentum conservation
of any localized material system is violated
in a nonconstant gravitational background.
The impact for experiment appears, however, negligible
if the vacuum-energy mass scale is of order $\text{meV}$.
\vspace*{1\baselineskip}\newline
\textit{Journal}: Int. J. Mod. Phys. D 26 (2017) 1750006 
\vspace*{.75\baselineskip}\newline
\textit{Preprint}:  arXiv:1604.03065  
\end{abstract}
\keywords{General relativity; dark energy; cosmology; cosmological constant.}


\section{Introduction}
\label{sec:Intro}

A novel point of view on the long-standing
cosmological constant problem
is provided by so-called unimodular
gravity.\cite{vanderBijvanDamNg1981,Zee1983,BuchmullerDragon1988,%
HenneauxTeitelboim1989,Weinberg1988}
In this approach, the cosmological constant $\Lambda$
does not appear as input of the gravitational field equation
but arises as a constant of integration.
However, unimodular gravity does not provide the
value of $\Lambda$.

Here, we present a theory which interpolates between standard
general relativity and unimodular gravity.
The difference with the standard formulation of unimodular gravity
is that a value for $\Lambda$ is built in, namely, $\Lambda=0$.
Specifically, we find that the solution of the
gravitational field equation in a cosmological context
approaches a metric with determinant minus 1 and has a
vanishing cosmological constant (vacuum energy density).

Still, the new theory does not have the main advantage of the
unimodular-gravity approach, the automatic cancellation of
zero-point energies. These zero-point energies and other
contributions to the vacuum energy density are to be cancelled
dynamically by microscopic degrees of freedom\cite{Volovik2008}
and this cancellation can be described macroscopically by the so-called
$q$-theory.\cite{KlinkhamerVolovik2008a,KlinkhamerVolovik2008b,%
KlinkhamerVolovik2010,KlinkhamerVolovik2016a,KlinkhamerVolovik2016b}
The present article assumes that
$q$-theory (or any other viable compensation
mechanism) provides for a cancellation of
the  zero-point energies
and focusses on the gravitation theory emerging in the infrared.

Our attitude as to which type of
gravitation theory emerges is entirely agnostic
and, in principle, we wish to explore all
possible terms involving the metric field
(the gravitational field definitely known to exist).
For a theory valid over large distances, we then
look for terms with the lowest number of derivatives
of the metric. It turns out that, apart from the term
corresponding to the cosmological constant $\Lambda$,
there is another term without derivatives.
This term involves the determinant of the metric,
hence the connection to unimodular gravity.
But there is a price to pay if we wish to maintain
general covariance, as will be explained below.

\section{Generalized unimodular gravity}
\label{sec:Generalized-unimodular-gravity}

\subsection{Setup}
\label{subsec:Setup}

Consider a modest extension of standard general
relativity with the following action:%
\bsubeqs\label{eq:Action-g-sigma}
\beqa
\label{eq:Action}
 S&=&\
S^\text{\,grav} + S^{\,M}
 =
- \int_{\mathbb{R}^4}
\,d^4x\, \sqrt{-g}\,\left(\frac{R}{16\pi G_{N}}
+\epsilon(\sigma)+\mathcal{L}^{\,M}[\psi]\right) \,,
\\[2mm]
\label{eq:g}
g &\equiv& \det g_{\alpha\beta} \,,
\\[2mm]
\label{eq:sigma}
\sigma &\equiv& \sqrt{-g}/\sqrt{-g_{0}} \,,
\eeqa
\esubeqs
where the metric $g_{\alpha\beta}$ has a Lorentzian signature
($-,\,+,\,+,\,+$) and where
$g_{0} = g_\text{Mink}$ corresponds to the determinant of the 
Minkowski metric $g^\text{Mink}_{\alpha\beta}$ with the same signature.
Remark that $\sigma$ is a scalar, as it is
the ratio of two scalar densities of equal weight.
The theory \eqref{eq:Action-g-sigma}
is, in fact, invariant under  general coordinate  transformations  with
arbitrary nonzero values of
$\det(\partial x'^{\,\alpha}/\partial x^{\beta})$.
The price to pay for having unrestricted general coordinate invariance
is the introduction of a prior metric giving the scalar density
$g_{0}$. As we will see in the next subsection, this
price is relatively small, just a ``parameter'' of a potential
term in the action.

The field $\psi$ in the Lagrange density $\mathcal{L}^{\,M}$
of the matter action $S^{\,M}$
stands for a generic matter field containing, for example,
all the fields of the standard model of elementary particle physics.
Strictly speaking,   $\mathcal{L}^{\,M}$ also depends on
the metric or the vierbeins via the covariant derivatives,
but we keep this dependence implicit. In principle,
it is also possible to make the gravitational coupling $G$ 
and the parameters of $\mathcal{L}^{\,M}$ dependent on $\sigma$,
that is, to have $G=G(\sigma)$ replacing Newton's constant $G_{N}$ 
and $\mathcal{L}^{\,M}=\mathcal{L}^{\,M}(\sigma,\psi)$. 
But, here, we consider the simplest possible theory
with action \eqref{eq:Action} and the scalar $\sigma$ appearing only in the
potential term $\epsilon(\sigma)$.

The Einstein gravitational field equation from \eqref{eq:Action}
takes the standard form,
\beq\label{eq:Einstein-eq}
R^{\alpha\beta} -\half\;g^{\alpha\beta}\;R
 =
 -8\pi G_{N}\left(T_{V}^{\alpha\beta}+T_{M}^{\alpha\beta}\right) \,,
\eeq
with the standard energy-momentum tensor of the matter fields,
\beq\label{eq:TM-alpha-beta}
T_{M}^{\alpha\beta}
\equiv
\frac{2}{\sqrt{-g}}\;
\frac{\delta S_{\,M}}{\delta g_{\alpha\beta}}\,,
\eeq
and the vacuum-energy term
\bsubeqs\label{eq:TalphabetaV-rhoV}
\beqa\label{eq:TalphabetaV}
 T_{V}^{\alpha\beta}&=& \rho_{V}(\sigma)\,g^{\alpha\beta} \,,
\\[2mm]
\label{eq:rhoV}
\rho_{V}(\sigma) &=&  \epsilon(\sigma) + \sigma\;\frac{d\epsilon(\sigma)}{d\sigma}\,.
\eeqa
\esubeqs
The crucial observation here is that the vacuum energy density $\rho_{V}$
of the gravitational field equation differs from the
vacuum energy density $\epsilon$ of the action.
A similar difference has been found before in the
context of condensed matter physics\cite{Volovik2008}  and certain
relativistic theories.\cite{KlinkhamerVolovik2008a,KlinkhamerVolovik2008b,%
KlinkhamerVolovik2010}

For concreteness, take the following quadratic
\textit{Ansatz} for the function
$\epsilon(\sigma)$, with corresponding $\rho_{V}(\sigma)$
from \eqref{eq:rhoV}:
\bsubeqs\label{eq:epsilonAnsatz2-rhoVAnsatz2}
\beqa\label{eq:epsilonAnsatz2}
\epsilon(\sigma) &=& m^4\,
\left[\frac{1}{3}\;\left(\sigma-\frac{3}{2} \right)^2 + \frac{1}{4} \right] \,,
\\[2mm]
\label{eq:rhoVAnsatz2}
\rho_{V}(\sigma) &=&  m^4\,\left[1-\sigma \right]^2  \,,
\eeqa
\esubeqs
where  $m$ is a new mass scale.
Admittedly, we have fine-tuned \eqref{eq:epsilonAnsatz2},
in order to arrive at \eqref{eq:rhoVAnsatz2}.
See Sec.~\ref{subsec:Possible-underlying-physics}
for discussion on how
the underlying physics, perhaps analogous to known
condensed-matter-physics systems, could produce
\eqref{eq:rhoVAnsatz2} close to equilibrium.

The theory given by Eqs.~\eqref{eq:Action-g-sigma}
and \eqref{eq:epsilonAnsatz2-rhoVAnsatz2}
can be considered as an interpolation between standard
general relativity and unimodular gravity.
The limit $m\to 0$ reproduces standard general relativity
and  the limit $m\to \infty$ gives unimodular gravity,
in the sense that $\det g_{\alpha\beta}(x)$
for standard Cartesian coordinates
is fixed dynamically to  a constant value.
As an effective theory, we may consider setting
$m \sim \text{meV}$, in line with astronomical observations of
the present accelerating universe.\cite{PDG2014}

In this article, we consider only a
prior metric corresponding to Minkowski spacetime,
but the afore-mentioned astronomical observations
suggest the relevance of a prior metric corresponding
to de-Sitter spacetime, which we briefly discuss
in \ref{app:Prior-metric-of-de-Sitter-spacetime}.

\subsection{Equilibrium conditions and linearized gravity}
\label{subsec:Equilibrium-conditions-linearized-gravity}

The \textit{Ansatz} \eqref{eq:epsilonAnsatz2-rhoVAnsatz2}
implements the following equilibrium conditions
at $\sigma_{0}=1$:
\bsubeqs\label{eq:equilibrium-conditions-a-b-c}
\beqa\label{eq:equilibrium-conditions-a}
\rho_{V}(\sigma_{0}) &=& 0\,,
\\[2mm]
\label{eq:equilibrium-conditions-b}
\left[\frac{d\rho_{V}(\sigma)}{d\sigma}\right]_{\sigma_{0}} &=& 0\,,
\\[2mm]
\label{eq:equilibrium-conditions-c}
\left[\frac{d^2\rho_{V}(\sigma)}{d\sigma^2}\right]_{\sigma_{0}} &>& 0\,.
\eeqa
\esubeqs
In this way, we have a generalization of unimodular gravity,
given by Eqs.~\eqref{eq:Action-g-sigma} and \eqref{eq:epsilonAnsatz2-rhoVAnsatz2},
which reproduces the linearized
theory of general relativity (here, in the harmonic gauge)
but differs in higher order:
\bsubeqs\label{eq:Ansatz2-metric-perturbation-harmonic-gauge-Box-h-linearized}
\beqa
\label{eq:Ansatz2-metric-perturbation}
g_{\alpha\beta} &=& \eta_{\alpha\beta}+  h_{\alpha\beta}\,,
\\[2mm]
\label{eq:Ansatz2-harmonic-gauge}
\partial_{\alpha}\, h^{\alpha}_{\;\;\beta}  &=&
\frac{1}{2}\, \partial_{\beta}\, h^{\alpha}_{\;\;\alpha} \,,
\\[2mm]
\label{eq:Ansatz2-Box-h-linearized}
\Box \, h_{\alpha\beta}
&=&
-16\pi G_{N}\,\left[ \widehat{T}_{\alpha\beta}^{\,M}
- \frac{1}{2}\,\eta_{\alpha\beta}\,
\widehat{T}^{\,M\,\gamma}_{\phantom{M}\;\;\;\gamma}\right]
+ \cdots\,,
\eeqa
\esubeqs
with the Minkowski metric
$\eta_{\alpha\beta}\equiv [\text{diag}(-1,1,1,1)]_{\alpha\beta}$
for standard Cartesian coordinates,
the flat-spacetime d'Alembertian $\Box \equiv \partial_0^2 -\nabla^2$,
and the flat-spacetime energy-momentum tensor
$\widehat{T}_{\alpha\beta}^{\,M}$ containing only $\eta_{\alpha\beta}$.

The  equilibrium conditions \eqref{eq:equilibrium-conditions-a-b-c}
make that the contribution \eqref{eq:TalphabetaV}
to the gravitational field equation
is second-order in $(\sigma_{0}-\sigma)$ and
corresponds to quartic order in $h$ after gauge fixing (see below).
Strictly speaking,
the condition \eqref{eq:equilibrium-conditions-a}
suffices to recover the linearized
theory of general relativity, but we add the
equilibrium condition \eqref{eq:equilibrium-conditions-b}
and the further stability condition \eqref{eq:equilibrium-conditions-c}.
These conditions arise naturally in the framework 
of $q$-theory\cite{KlinkhamerVolovik2008a,%
KlinkhamerVolovik2010,KlinkhamerVolovik2016b}:
conditions \eqref{eq:equilibrium-conditions-a}
and \eqref{eq:equilibrium-conditions-b} come from the
self-adjustment of the conserved vacuum variable $q$ [with chemical
potential $\mu=d\epsilon/dq$ and $\rho_{V}(q) =  \epsilon(q) -\mu\,q$]
by use of the Gibbs--Duhem relation for an isolated
self-sustained system without external pressure,
while condition \eqref{eq:equilibrium-conditions-c}
corresponds to having a positive isothermal compressibility.
See Sec.~\ref{subsec:Possible-underlying-physics} for further
discussion.

For gravitational waves in the transverse-traceless
(TT) gauge (see, e.g., Sec. 35.4 of Ref.~\refcite{MTW1973}),
we obtain from
\eqref{eq:Ansatz2-Box-h-linearized}
without source terms the standard linear wave equation:
\beq\label{eq:standard-linear-wave-equation-TT}
\Box \, h_{\alpha\beta}^{TT} = 0\,,
\eeq
which gives standard propagation,
\beq\label{eq:standard-propagation}
k^2=\eta^{\alpha\beta}\,k_{\alpha}\,k_{\beta} =0  \,.
\eeq
The standard propagation behavior \eqref{eq:standard-propagation}
differs from the generic propagation behavior of, for example,  
Rosen's bi-metric theory discussed in Ref.~\refcite{Will1993}.
The crux is that, for us, the prior metric $\eta_{\alpha\beta}$
enters only in the potential term $\epsilon$
of \eqref{eq:Action} and not in the kinetic terms
$R$ and $\mathcal{L}^{\,M}$. In this way, the d'Alembertian
for gravitational  waves is the same as the one for
electromagnetic waves from the Maxwell term
$F_{\alpha\beta}\,F^{\alpha\beta}$ contained in $\mathcal{L}^{\,M}$.

In closing, we display the scalar
$\sigma$ in terms of the physical degrees of freedom.
Consider a plane gravitational wave in the TT gauge
propagating in the 3-direction and denote the two
polarizations by $h_{+}$ and $h_{\times}$.
The $2\times 2$ sub-matrix for $h_{\alpha\beta}$ has $\pm h_{+}$
on the diagonal positions and $h_{\times}$ on the off-diagonal positions.
We then have the determinant
\beq
\det g_{\alpha\beta} = -\big(1- h_{+}^2-h_{\times}^2 \big)\,,
\eeq
so that  
\beq
\sigma = 1- \half\,( h_{+}^2 + h_{\times}^2)\,,
\eeq
for $|h_{+}| \ll 1$ and $|h_{\times}| \ll 1$.
The \textit{Ansatz} \eqref{eq:rhoVAnsatz2} now gives
$\rho_{V}(\sigma)$  $\propto$ $( h_{+}^2+h_{\times}^2)^2$,
which does not affect the standard result
\eqref{eq:standard-linear-wave-equation-TT}.

\subsection{Vacuum-matter energy exchange from a local action}
\label{subsec:Vacuum-matter-energy-exchange-from-a-local-action}

For homogeneous matter fields in a cosmological context,
it can be shown that the vanishing covariant divergence of
Eq.~\eqref{eq:TalphabetaV} gives $\partial_t\,\rho_V = 0$.
A nonconstant $\rho_V$ apparently requires energy exchange
between the vacuum energy density and the matter energy density,
but this energy exchange
is not present in the basic theory \eqref{eq:Action-g-sigma}.
The simplest possible way to implement
this energy exchange is to change the matter action,
so that the matter component by itself
does not conserve energy-momentum.

Consider, then, a
real scalar field $\phi(x)$ with a non-dynamical dimensionless
real scalar field $\zeta(x)$ in the mass-square term ($\hbar=c=1$):
\beq\label{eq:action-phi-zeta}
\overline{S}^{\,M} =
 - \int_{\mathbb{R}^4} \,d^4x\, \sqrt{-g}\;
\Big(\half\,g^{\alpha\beta}\,\partial_{\alpha}\phi\,\partial_{\beta}\phi
+\half\,\Big[M^2 + m^2\,\zeta\Big]\,\phi^2 \Big) \,,
\eeq
in terms of the mass-scale $m$ of the
vacuum energy density \eqref{eq:epsilonAnsatz2-rhoVAnsatz2}.
The corresponding energy-momentum tensor is obtained from
\beq\label{eq:Tbar-alpha-beta}
\overline{T}_{M}^{\alpha\beta}
\equiv
\frac{2}{\sqrt{-g}}\;
\frac{\delta \overline{S}_{\,M}}{\delta g_{\alpha\beta}}\,.
\eeq
Using the field equation of $\phi$, the energy-momentum tensor
\eqref{eq:Tbar-alpha-beta}
turns out to have a nonvanishing covariant divergence
for nonconstant $\zeta(x)$,
\beq\label{eq:Tbar-nonconservation}
\nabla_{\alpha}\overline{T}_{M}^{\alpha\beta}
=-\half\,m^2\,\phi^2\:\partial^{\beta}\,\zeta\,.
\eeq
The reason for this nonconservation is that there is no
field equation for $\zeta(x)$; see, for example, the discussion
around Eq.~(E.1.27) in Ref.~\refcite{Wald1984}
where the generic matter field $\psi$ is
replaced by our scalar fields $\phi(x)$ and $\zeta(x)$.
Still, there is total energy-momentum conservation,
\beq\label{eq:energy-momentum-conservation-general}
\nabla_{\alpha}
\left(T_{V}^{\alpha\beta}+\overline{T}_{M}^{\alpha\beta}\right)=0\,,
\eeq
which follows from the contracted Bianchi
identities\cite{MTW1973,Wald1984}
and the Einstein gravitational field equation \eqref{eq:Einstein-eq}
with $T_{V}^{\alpha\beta}$ from \eqref{eq:TalphabetaV-rhoV}
and $\overline{T}_{M}^{\alpha\beta}$  from \eqref{eq:Tbar-alpha-beta}.
In fact,
$\zeta(x)$ is determined by \eqref{eq:energy-momentum-conservation-general},
as shown by the explicit example in
Sec.~\ref{subsec:Nonstatic-universe-from-vacuum-matter-energy-exchange}.
Observe that
Eq.~\eqref{eq:Tbar-nonconservation} is time-reversal invariant
and corresponds to a nondissipative process, different from
the one considered in Ref.~\refcite{KlinkhamerVolovik2016a}.

The modified energy-momentum conservation of the matter
\eqref{eq:Tbar-nonconservation} can be expected to affect
the generation of gravitational waves,\cite{PoissonWill2014}
but this topic lies outside the scope of the present paper.
Possible experimental consequences of the
modified energy-momentum conservation are, however, briefly
discussed in Sec.~\ref{subsec:Experiments}.

\section{Cosmology}
\label{sec:Cosmology}

\subsection{Metric}
\label{subsec:Metric}

Take the particular  generalization of unimodular gravity
given by Eqs.~\eqref{eq:Action-g-sigma},
\eqref{eq:epsilonAnsatz2-rhoVAnsatz2}, and \eqref{eq:action-phi-zeta}.
Consider, now, the
spatially-flat ($k=0$) Robertson--Walker (RW)
metric for standard comoving Cartesian coordinates
and for rescaled spatial coordinates with the line element given by
\bsubeqs\label{eq:flat-RW-metric-a-b-b-definition}
\beqa\label{eq:flat-RW-metric-a}
ds^2 &=&-dt^2 + a^2(t)\,
\left[(dx^{1})^2+(dx^{2})^2+(dx^{3})^2\right] \,,
\\[2mm]
\label{eq:flat-RW-metric-b}
&=&-dt^2 + b^2(t)\,
\left[(d\widetilde{x}^{1})^2+(d\widetilde{x}^{2})^2+(d\widetilde{x}^{3})^2\right] \,,
\\[2mm]
\label{eq:flat-RW-metric-b-definition}
b(t) &=& \frac{a(t)}{a(t)+1}\,,
\eeqa
\esubeqs
where the bounded expansion factor $b(t)$ will be used from now on.
Minkowski spacetime in standard Cartesian coordinates
has the metric \eqref{eq:flat-RW-metric-b}
with
\beq\label{eq:flat-RW-metric-b-Minkowski}
b(t) = 1 \,.
\eeq
Further rescaling of the spatial coordinates  in Minkowski spacetime
can, of course,  give any positive constant value for $b$.

\subsection{Nonstatic universe from vacuum-matter energy exchange}
\label{subsec:Nonstatic-universe-from-vacuum-matter-energy-exchange}

For the Robertson--Walker metric \eqref{eq:flat-RW-metric-b}
with homogeneous matter fields $\phi(t)$
and $\zeta(t)$, we get from \eqref{eq:Action-g-sigma},
\eqref{eq:epsilonAnsatz2-rhoVAnsatz2}, and \eqref{eq:action-phi-zeta}
the Klein--Gordon equation and the two Friedmann equations:%
\bsubeqs\label{eq:Friedmann-eqs-b}
\beqa
\label{eq:Friedmann-eqs-KG-b}
\partial_{t}^2\,\phi  + 3\,H\,\partial_{t}\,\phi + M^2 \,\phi
&=&
- m^2\,\zeta\,\phi\,,
\\[2mm]
\label{eq:Friedmann-eqs-H2-b}
3\,H^2 &=&
 8\pi G_{N} \, \big(\rho_{V}+\rho_{M}\big)\,,
\\[2mm]
\label{eq:Friedmann-eqs-dotH-b}
2\,\,\partial_{t}\,H &=& -8\pi G_{N}\, \big(\rho_{M}+ P_{M}\big) \,,
\eeqa
with the Hubble parameter,
the matter energy density, the matter pressure, and
the vacuum energy density given by
\beqa
\label{eq:Friedmann-eqs-H-b}
H&=& \frac{\partial_{t}\, b}{b\,(1-b)} \,,
\\[2mm]
\label{eq:rho-M-eq-b}
\rho_{M} &=&
\half\,(\partial_{t}\,\phi)^2 + \half\,\big[M^2 + m^2\,\zeta\big]\,\phi^2\,,
\\[2mm]
\label{eq:P-M-eq-b}
P_{M} &=&
\half\,(\partial_{t}\,\phi)^2 - \half\,\big[M^2 + m^2\,\zeta\big]\,\phi^2\,,
\\[2mm]
\label{eq:rhoV-b}
\rho_{V} &=& -P_{V} = m^{4}\,\left[1-b^3\,\right]^2 \,.
\eeqa
\esubeqs

Turning to the issue of vacuum-matter energy exchange,  start with the time derivative
of the first-Friedmann Eq.~\eqref{eq:Friedmann-eqs-H2-b},
\beq
\label{eq:derivative-Friedmann-eqs-H2-b}
6\,H\,\partial_{t}\,H
=
8\pi G_{N} \, \big(\partial_{t}\,\rho_{V}+\partial_{t}\,\rho_{M}\big)\,.
\eeq
The left-hand-side of \eqref{eq:derivative-Friedmann-eqs-H2-b}
can be eliminated by use of $3 H$ times
\eqref{eq:Friedmann-eqs-dotH-b} to give
the expression of total energy-momentum conservation,
\beq\label{eq:energy-momentum conservation}
\partial_{t}\,\rho_{M}  + 3\,H\, \big(\rho_{M}+ P_{M}\big)
= - \partial_{t}\,\rho_{V}\,.
\eeq
Now, the left-hand-side of
\eqref{eq:energy-momentum conservation}
can be evaluated explicitly by use of the
definitions \eqref{eq:rho-M-eq-b}
and \eqref{eq:P-M-eq-b} and the
Klein--Gordon equation \eqref{eq:Friedmann-eqs-KG-b}.
The result is
\bsubeqs
\beqa
\label{eq:rho-M-dot-eq-b}
\partial_{t}\,\rho_{M}  + 3\,H\, \big(\rho_{M}+ P_{M}\big)
&=&
\half\,m^2\,\phi^2\,\partial_{t}\,\zeta\,,
\eeqa
which, together with \eqref{eq:energy-momentum conservation}, gives
\beqa
\label{eq:rho-V-dot-eq-b}
\partial_{t}\,\rho_{V}
&=&
-\half\,m^2\,\phi^2\,\partial_{t}\,\zeta\,.
\eeqa
\esubeqs
Result \eqref{eq:rho-M-dot-eq-b}
also follows directly from \eqref{eq:Tbar-nonconservation}.

For constant $\zeta$,  i.e.
without essential modification of the mass-square term in
the scalar action \eqref{eq:action-phi-zeta},
the result \eqref{eq:rho-V-dot-eq-b}
shows that $\partial_{t}\,\rho_{V}$ vanishes
and from \eqref{eq:rhoV-b}
we then have a static universe with $b(t)=\text{const}$.
For $\phi^2\,\partial_{t}\,\zeta \ne 0$,
the source terms on the right-hand-sides
of  \eqref{eq:rho-M-dot-eq-b}
and \eqref{eq:rho-V-dot-eq-b}
show the nonvanishing energy exchange and allow for
$\partial_{t}\,\rho_{V}\ne 0$.
Equations~\eqref{eq:rhoV-b} and \eqref{eq:rho-V-dot-eq-b}
then imply that $\phi$ must be of order $m$, as long as
$\zeta$ is of order unity.

\subsection{Numerical results}
\label{subsec:Numerical-results}

The mass-scale $m$ of the vacuum energy density
and the Planck energy $E_{P} \equiv
G_{N}^{-1/2} \approx 1.22 \times 10^{19}\,\text{GeV}$
give the following mass-square ratio:
\beq\label{eq:xi-def}
\xi \equiv (m/E_{P})^2  \,.
\eeq
Next, use $E_{P}$ to define the dimensionless time $\tau$,
the dimensionless Hubble parameter $h(\tau)$,
the dimensionless scalar field  $\varphi(\tau)$,
and the dimensionless
vacuum energy density $r_{V}(\tau)=\xi^2\,[b(\tau)^3-1]^2$.
Henceforth, an overdot will denote differentiation with respect
to $\tau$.  For an initial study, we take all masses equal to the
Planck scale, $M=m=E_{P}$.

\begin{figure*}[t]  
\vspace*{8mm}
\begin{center}   
\includegraphics[width=\textwidth]
{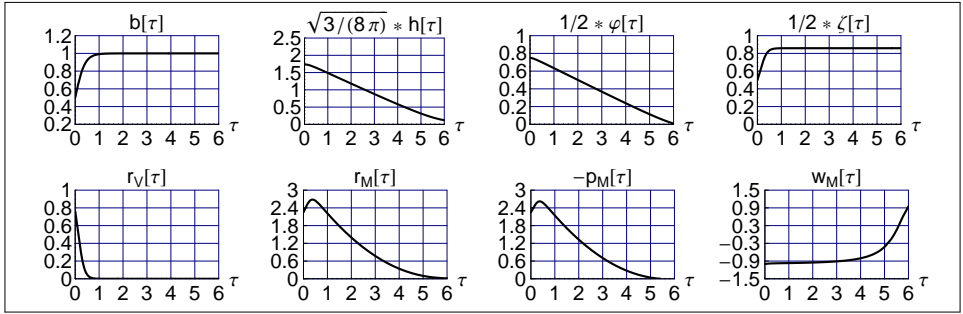}
\end{center}   
\vspace*{1mm}
\caption{Numerical solutions $b(\tau)$, $\varphi(\tau)$, and $\zeta(\tau)$
from the ODEs \eqref{eq:Friedmann-eqs-KG-b},
\eqref{eq:Friedmann-eqs-dotH-b},
and \eqref{eq:derivative-Friedmann-eqs-H2-b},
with further definitions \eqref{eq:Friedmann-eqs-H-b},
\eqref{eq:rho-M-eq-b}, \eqref{eq:P-M-eq-b},
and \eqref{eq:rhoV-b}.
Also shown are the following derived quantities:
the Hubble parameter $h$, the vacuum energy density $r_{V}$,
the  matter energy density $r_{M}$,
the  matter pressure $p_{M}$, and
the  matter equation-of-state parameter $w_{M}\equiv p_{M}/r_{M}$.
All masses have been taken equal to the Planck mass and the corresponding
model parameter is $\xi=1$.
The boundary conditions at $\tau=0$ are
$\{b(0),\,\dot{b}(0),\,\phi(0),\,\dot{\phi}(0),\,\zeta(0)\}$
$=$ $\{ 1/2,\,1.25657,\,3/2,\,0,\,1\}$.
The vacuum energy density $r_{V}(\tau) =\xi^2\,[1-b(\tau)^3\,]^2$ takes
the values
$r_V(0)\approx 0.77$  and $r_V(6)\approx 3 \times 10^{-13}$.
}
\label{fig:flat-FRW-b-phi-zeta}
\end{figure*}

Numerical results are obtained by solving the Klein--Gordon
Eq.~\eqref{eq:Friedmann-eqs-KG-b},
the second-Friedmann Eq.~\eqref{eq:Friedmann-eqs-dotH-b},
and the time derivative of
the first-Friedmann Eq.~\eqref{eq:Friedmann-eqs-H2-b}.
This last equation, given as
Eq.~\eqref{eq:derivative-Friedmann-eqs-H2-b},
contains the derivatives
$\ddot{b}$, $\dot{b}$, $\ddot{\varphi}$, $\dot{\varphi}$,
and, in particular, $\dot{\zeta}$.
The boundary conditions at $\tau=0$
are $\{b(0),\,\phi(0),\,\dot{\phi}(0),\,\zeta(0)\}$
with $\dot{b}(0)$ determined by
the first-Friedmann Eq.~\eqref{eq:Friedmann-eqs-H2-b}.
The numerical results of Fig.~\ref{fig:flat-FRW-b-phi-zeta}
show that Minkowski spacetime is approached with
$b\to 1$, $h\to 0$, $\phi\to 0$, and $\zeta \to \text{const}$.

Essentially the same results as in Fig.~\ref{fig:flat-FRW-b-phi-zeta}
are obtained for the following initial conditions:
\beq
\{b(0),\,\phi(0),\,\dot{\phi}(0),\,\zeta(0)\}
=
\{1/2 \pm 1/10,\,3/2 \pm 1/5,\,\pm 1/100,\,1\pm 1/100\}\,,
\eeq
with corresponding $\dot{b}(0)$ values from
the first-Friedmann Eq.~\eqref{eq:Friedmann-eqs-H2-b}.
This establishes numerically the Minkowski attractor behavior for
a finite domain of initial conditions.

For model parameter $\xi < 1$,
the cosmic time unit of Fig.~\ref{fig:flat-FRW-b-phi-zeta}
is scaled up with  a factor $1/\xi$, which
takes the approximate numerical value $1/\xi\sim 10^{62}$
for $m\sim \text{meV}$. As the time unit in
Fig.~\ref{fig:flat-FRW-b-phi-zeta}
is the Planck time $t_{P}$, the rescaled time unit becomes
$t_{P}/\xi \sim 10^{19}\;\text{s}$, which is
relatively close (no surprise)
to the inferred age of the present universe
$13 \times 10^{9}\;\text{yr}  \sim 4 \times 10^{17}\;\text{s}$.
Of course, the aim of the present work is not to
give an accurate description of the actual (accelerating) universe
but is more modest, namely, to investigate the
gravitating vacuum energy without the introduction of new fields.
Still, we have performed an exploratory calculation for
a prior metric corresponding to de-Sitter spacetime;
see \ref{app:Prior-metric-of-de-Sitter-spacetime} for details.

The model based on  Eqs.~\eqref{eq:Action-g-sigma},
\eqref{eq:epsilonAnsatz2-rhoVAnsatz2}
and \eqref{eq:action-phi-zeta} suffices as an effective gravity theory.
The ultimate question is, of course, the microscopic origin
of the non-dynamical field $\zeta(x)$. Some remarks are presented in
Sec.~\ref{subsec:Possible-underlying-physics}.

\section{Discussion}
\label{sec:Discussion}

\subsection{Experiments}
\label{subsec:Experiments}

The focus of this paper has been on cosmology,
but there may also be implications for small-scale
experiments.
Energy-momentum conservation of any localized material
system would be violated in the following way:
\beq\label{eq:matter-energy-momentum-violation}
\nabla_{\alpha}\,T_{M}^{\alpha\beta}
=
-\nabla_{\alpha}\,T_{V}^{\alpha\beta}
\stackrel{?}{=}
- \partial^{\beta}\,\rho_{V}(\sqrt{-g}/\sqrt{-g_\text{Mink}})
\stackrel{??}{=}
- m^4\;\partial^{\beta}\,[\sqrt{-g}/\sqrt{-g_\text{Mink}}-1]^2\,,
\eeq
with the definition $g \equiv \det g_{\alpha\beta}$ and
the suffix ``Mink'' standing for the prior metric
of Minkowski spacetime.
The equality with a single question mark
in \eqref{eq:matter-energy-momentum-violation}
assumes the validity of our effective theory \eqref{eq:Action-g-sigma}
with energy exchange from \eqref{eq:action-phi-zeta} and the equality with two question marks follows from
the $\rho_{V}$ \textit{Ansatz} \eqref{eq:rhoVAnsatz2}.

Let us get some orders of magnitude for two hypothetical experiments.
We know that,
at a large distance $R$ from a localized mass $M$, the order of
magnitude for $\sqrt{-g}/\sqrt{-g_\text{Mink}}-1$ is $O(G_{N}M/R)$,
and we are primarily interested in spatial derivatives of the
Newtonian potential. This gives the following order of
magnitude for the right-hand-side of \eqref{eq:matter-energy-momentum-violation}:
\beq\label{eq:matter-energy-momentum-violation-right-hand-side}
O(m^4\;\partial^{\beta}\,[\sqrt{-g}/\sqrt{-g_\text{Mink}}-1]^2)
=
O(m^4\,R^{-1}\,[G_{N}M/R]^2)
\,.
\eeq
For a laboratory experiment on Earth, the Newtonian potential is
$|\phi_N| = G_{N}M/R \sim 10^{-9}$ at $R \sim 10^{7}\;\text{m}$.
Consider, now, a quark-gluon-plasma (QGP)\cite{McLerran1986} possibly created
by heavy-ion collisions at CERN's Large Hadron Collider
(ALICE detector)
with $\rho_M \sim \text{GeV}^4$ and $L_M \sim  \text{fm} = 10^{-15}\;\text{m}$.
From \eqref{eq:matter-energy-momentum-violation} and
\eqref{eq:matter-energy-momentum-violation-right-hand-side},
the relative matter energy-momentum violation (rMEMV)
then  has the following order of magnitude:
\beqa\label{eq:rMEMV-QGP}
\hspace*{-11mm}&&
\text{rMEMV}\big|^\text{(QGP)}
\sim
\frac{m^4}{\rho_M}\;\frac{L_M}{R}\;\big(\phi_N\big)^2
\nonumber\\[2mm]
\hspace*{-11mm}&&
\sim
10^{-88}   
\;
\left(\frac{m^4}{10^{-12}\;\text{eV}^4}\right)\,
\left(\frac{10^{36}\;\text{eV}^4}{\rho_M}\right)\,
\left(\frac{L_M}{10^{-15}\;\text{m}}\right)\,
\left(\frac{10^{7}\;\text{m}}{R}\right)\,
\left(\frac{\phi_N}{10^{-9}}\right)^2,
\eeqa
for a vacuum-energy mass scale $m \sim \text{meV}$ 
as indicated by the present accelerating universe.\cite{PDG2014}

A different ``experiment'' concerns the binary black-hole merger (BBHM)
observed by LIGO.\cite{LIGO2016a}
Here, the metric perturbations are of order 1
and the length scale involved is of the order of the Schwarzschild radius
of the smaller initial black hole,
$R\sim 2\,G_{N}\,M_{BH} \sim 60\,G_{N}\, M_\text{Sun}\sim 45 \;\text{km}$.
The effective matter energy density is of order
$\rho_M \sim M_{BH}/(G_{N}M_{BH})^3  \sim E_{P}^6/(M_{BH})^2
\sim 10^{33}\;\text{eV}^4$ for $M_{BH} \sim 30\, M_\text{Sun}$.
All in all, we have the following order of magnitude for
the relative matter energy-momentum violation:
\bsubeqs\label{eq:rMEMV-BBHM-BNSM}
\beqa\label{eq:rMEMV-BBHM}
\hspace*{-0mm}
\text{rMEMV}\big|^\text{(BBHM)}
&\sim&
\frac{m^4}{\rho_M}
\sim
10^{-45}  
\;\left(\frac{m^4}{10^{-12}\;\text{eV}^4}\right)\,
\left(\frac{10^{33}\;\text{eV}^4}{\rho_M}\right)\,.
\eeqa
A  similar result is obtained for the (not yet observed)
coalescence of two neutron stars. 
This neutron-star result is perhaps more reliable
than the black-hole result \eqref{eq:rMEMV-BBHM},
because the matter energy density of the neutron stars
can be identified directly.
The result for a binary neutron-star merger (BNSM) is then
\beqa\label{eq:rMEMV-BNSM}
\hspace*{-0mm}
\text{rMEMV}\big|^\text{(BNSM)}
&\sim&
\frac{m^4}{\rho_M}\;\big(\phi_N\big)^2
\nonumber\\[2mm]&\sim&
10^{-47}
\;\left(\frac{m^4}{10^{-12}\;\text{eV}^4}\right)\,
\left(\frac{10^{33}\;\text{eV}^4}{\rho_M}\right)\,
\left(\frac{\phi_N}{10^{-1}}\right)^2\,.
\eeqa
\esubeqs

Provided the vacuum-energy mass scale is small
enough, $m \lesssim 10^{7}\;\text{eV}$,
the orders of magnitude \eqref{eq:rMEMV-QGP}
and \eqref{eq:rMEMV-BBHM-BNSM}
are extremely small. The predicted violation of matter
energy-momentum conservation from \eqref{eq:matter-energy-momentum-violation}
would then be negligible.
Generally speaking, the matter-energy-momentum-violating effect from
\eqref{eq:matter-energy-momentum-violation} would be largest for
a low-density system in a strong gravitational background,
with linear dimensions (length and time)
of the system not very much smaller than those of the background.

\subsection{Possible underlying physics}
\label{subsec:Possible-underlying-physics}

Condensed matter physics with emergent
topologically-protected Weyl fermions
(cf. Sec.~7.3 of Ref.~\refcite{Volovik2008} for a general discussion
and Refs.~\refcite{Foster-etal-2013,Foster-etal-2014}
for a detailed analysis in 2+1 dimensions)
demonstrates that known microscopic physics provides
the following ingredients:
\begin{enumerate}
\item[(i)]
formation of Weyl fermions obeying an effective gravity
with metric $\widetilde{g}_{\alpha\beta}$,
\item[(ii)]
nullification of the cosmological constant (vacuum energy density)
in full equilibrium,
\item[(iii)]
dynamical relaxation to the equilibrium Minkowski vacuum with or
without oscillations of the order parameter,
\item[(iv)]
mechanisms for the dissipative energy exchange between the coherent degrees of
freedom (vacuum) and the incoherent degrees of freedom (matter),
\item[(v)]
dependence of the vacuum energy on $\det \widetilde{g}_{\alpha\beta}$.
\end{enumerate}

The dynamics of a freely suspended two-dimensional 
film\cite{KatsLebedev2015} shows related effects for some of these items.
As shown in Ref.~\refcite{KlinkhamerVolovik2016b},
the theory of a two-dimensional film (2D brane)    
can be generalized to a (3+1)-dimensional theory (4D brane)
with gravity and an effective $q$ field of mass dimension 4.
For low-energy gravitational processes, this 4D brane theory
reproduces the action \eqref{eq:Action-g-sigma} with the 
prior metric corresponding to a constant number density on the brane.
The brane-type $q$-theory with energy scale $E_\text{br}$
has a quadratic vacuum energy density $\rho_{V}(q)$ near equilibrium, 
which,  in terms of the variable $\sigma$ from \eqref{eq:sigma}
with $\sigma \propto 1/q$, gives the expression 
$\rho_{V}(\sigma)=(E_\text{br})^4\,\sigma^{-2}\,\left[1-\sigma \right]^2$. 
The functional form of the latter expression 
is similar to the one of \textit{Ansatz} \eqref{eq:rhoVAnsatz2} used here.

In the present article, we have considered a generalization of unimodular
gravity which naturally follows from the ingredients listed above
and may come from a (3+1)-dimensional brane-type   
theory.\cite{KlinkhamerVolovik2016b}
The main result found here is that the vacuum-matter energy exchange is
uniquely determined by the \textit{Ansatz} for the vacuum energy density;
see, in particular, \eqref{eq:rho-V-dot-eq-b}
read from right (unknown) to left (known).
The reason for getting a prescribed source term of the matter
equation \eqref{eq:rho-M-dot-eq-b}
is that there is no new field entering the vacuum energy
density \eqref{eq:rhoV-b}.

But this uniquely determined vacuum-matter energy exchange
is surprising from a condensed-matter-physics point of view:
the energy exchange can be expected to depend
on the many details of the properties of matter,
and cannot be solely determined by the vacuum.
Perhaps the lesson for the underlying physics of gravity is that
\emph{if} the microscopic degrees of freedom really
give a vacuum energy density
$\rho_V$ as an effective function of the macroscopic metric
$g_{\alpha\beta}$ [for us, its determinant], then the microscopic
degrees of freedom \emph{also} arrange for the correct type of
energy exchange.

\subsection{Outlook}
\label{subsec:Outlook}

Without reliable information on the underlying physics of gravity,
we can take a more practical point of view.
The crucial input is Hubble's distance-redshift relation,
interpreted as coming from an expanding universe.
The present article has argued that,
if gravity over large distances has a vacuum component
$\rho_{V}(\sqrt{-g}/\sqrt{-g_\text{Mink}})$, then
a nonstatic universe requires that
there exists an energy exchange between this
vacuum component and the matter component.
The main task is to determine
(from observational cosmology or by laboratory experiments)
whether or not there is a contribution
$\rho_{V}(\sqrt{-g}/\sqrt{-g_\text{Mink}})$ to the vacuum energy density.

\section*{Acknowledgments}

It is a pleasure to thank G.E. Volovik for valuable discussions and comments.

\begin{appendix}
\section{Prior metric of de-Sitter spacetime}
\label{app:Prior-metric-of-de-Sitter-spacetime}

In the main text, we have used the prior metric of Minkowski
spacetime.
In this appendix, we take instead a prior metric corresponding
to de-Sitter spacetime with a positive cosmological constant,
\beq\label{eq:Lambda-app}
\Lambda = m^4 \,.
\eeq
This prior metric $g^\text{dS}_{\alpha\beta}$
is given by \eqref{eq:flat-RW-metric-b} with scale factor
\beq\label{eq:bdS-app}
b_\text{dS}(t)
=
\frac{\exp\left[\sqrt{(8\pi G_{N}/3)\,\Lambda}\;\;t\right]}
     {\exp\left[\sqrt{(8\pi G_{N}/3)\,\Lambda}\;\;t\right]-1+1/b_{\text{dS}0}}  \;,
\eeq
where the normalization parameter $b_{\text{dS}0}=b_{\text{dS}0}(\Lambda)$
has $b_{\text{dS}0}(0)=1$
and $b_{\text{dS}0}(\Lambda)<1$ for $\Lambda>0$.

\begin{figure*}[t]  
\vspace*{0mm}
\begin{center}    
\includegraphics[width=\textwidth]
{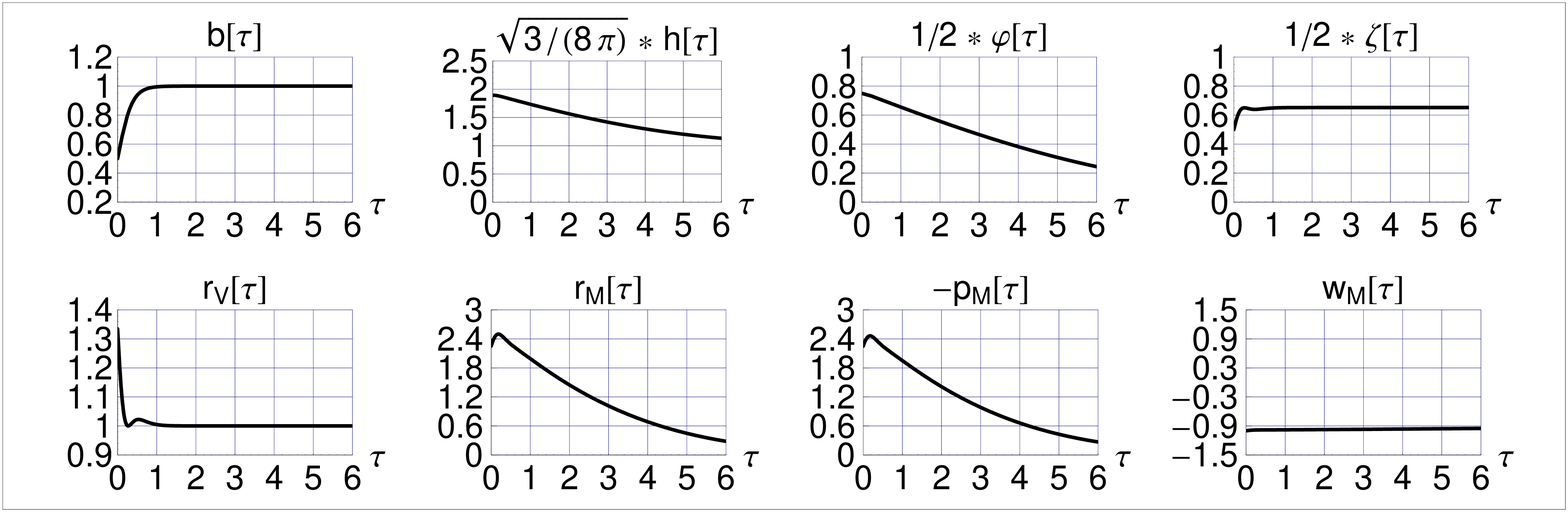}
\end{center}   
\vspace*{1mm}
\caption{Numerical solutions $b(\tau)$, $\varphi(\tau)$, and $\zeta(\tau)$
from the ODEs \eqref{eq:Friedmann-eqs-KG-b},
\eqref{eq:Friedmann-eqs-dotH-b},
and \eqref{eq:derivative-Friedmann-eqs-H2-b},
with further definitions \eqref{eq:Friedmann-eqs-H-b},
\eqref{eq:rho-M-eq-b}, \eqref{eq:P-M-eq-b},
and \eqref{eq:rV-app}, which includes the cosmological constant
$\Lambda=m^4$.
Also shown are the following derived quantities:
the Hubble parameter $h$, the vacuum energy density $r_{V}$,
the  matter energy density $r_{M}$,
the  matter pressure $p_{M}$, and
the  matter equation-of-state parameter $w_{M}\equiv p_{M}/r_{M}$.
All masses have been taken equal to the Planck mass and the
model parameters are $\xi=1$ and $b_{\text{dS}0}=2/3$.
The boundary conditions at $\tau=0$ are
$\{b(0),\,\dot{b}(0),\,\phi(0),\,\dot{\phi}(0),\,\zeta(0)\}$
$=$ $\{ 1/2,\,  1.36993,\,3/2,\,0,\,1\}$.
The vacuum energy density $r_{V}(\tau)$ takes
the values $r_V(0)\approx 1.33$  and $r_V(6)\approx 1$.
}
\label{fig:flat-FRW-b-phi-zeta-deSitter}
\end{figure*}

In the Lagrange density of \eqref{eq:Action-g-sigma} we have the potential term 
$\epsilon(\sigma)$,
for which we take the following \emph{Ansatz}
\bsubeqs\label{eq:epsilonAnsatz2-rhoVAnsatz2-sigma-app}
\beqa\label{eq:epsilonAnsatz2-app}
\epsilon(\sigma) &=& m^4\,
\left[\frac{1}{3}\;\left(\sigma-\frac{3}{2} \right)^2 + \frac{5}{4} \right] \,,
\\[2mm]
\label{eq:rhoVAnsatz2-app}
\rho_{V}(\sigma) &=&  m^4+m^4\,\left(1-\sigma \right)^2  \,,
\\[2mm]
\label{eq:sigma-app}
\sigma &=&  \sqrt{-g}/\sqrt{-g_\text{dS}}  \,.
\eeqa
\esubeqs
The matter action is still given by \eqref{eq:action-phi-zeta},
in order to have vacuum-matter energy exchange.

Turning to cosmology, we take the metric \eqref{eq:flat-RW-metric-b}
and consider
homogeneous matter fields $\phi(t)$ and $\zeta(t)$.
The dynamic equations are again given by \eqref{eq:Friedmann-eqs-KG-b},
\eqref{eq:Friedmann-eqs-H2-b},
and \eqref{eq:Friedmann-eqs-dotH-b}, with
definitions  \eqref{eq:Friedmann-eqs-H-b},
\eqref{eq:rho-M-eq-b}, and \eqref{eq:P-M-eq-b}.
In addition, there is now
the following vacuum energy density with cosmological constant
\eqref{eq:Lambda-app} included,
first in terms of dimensional variables and then
in terms of dimensionless variables:
\bsubeqs\label{eq:rhoV-rV-app}
\beqa
\label{eq:rhoV-app}
\rho_{V}(t) &=& -P_{V}(t)
= m^4\,\Big[1+ \Big(1-b(t)^3/b_\text{dS}(t)^3\Big)^2 \;\Big] \,,
\\[2mm]
\label{eq:rV-app}
r_{V}(\tau) &=& -p_{V}(\tau)
= \xi^2\,\Big[1+ \Big(1-b(\tau)^3/b_\text{dS}(\tau)^3\Big)^2 \;\Big] \,.
\eeqa
\esubeqs
Numerical results are shown in Fig.~\ref{fig:flat-FRW-b-phi-zeta-deSitter},
where the Hubble parameter $H(t)$ approaches the constant
value $\sqrt{(8\pi G_{N}/3)\,\Lambda}$
and the matter field $\phi(t)$ approaches $0$ while keeping
its equation-of-state parameter $w_{M}$ close to $-1$.

\end{appendix}


\end{document}